\documentclass[dvips]{article}
\usepackage{supertabular,lscape,epsfig}
\usepackage{amssymb}
\usepackage{amsmath}

\DeclareSymbolFont{ppa}{OT1}{ppl}{m}{it}
\DeclareMathSymbol{\vv}{\mathalpha}{ppa}{'166}

\begin{document}

\newcommand{\dd}{\,{\rm d}}
\newcommand{\ie}{{\it i.e.},\,}
\newcommand{\etal}{{\it et al.\ }}
\newcommand{\eg}{{\it e.g.},\,}
\newcommand{\cf}{{\it cf.\ }}
\newcommand{\vs}{{\it vs.\ }}
\newcommand{\zdot}{\makebox[0pt][l]{.}}
\newcommand{\up}[1]{\ifmmode^{\rm #1}\else$^{\rm #1}$\fi}
\newcommand{\dn}[1]{\ifmmode_{\rm #1}\else$_{\rm #1}$\fi}
\newcommand{\upd}{\up{d}}
\newcommand{\uph}{\up{h}}
\newcommand{\upm}{\up{m}}
\newcommand{\ups}{\up{s}}
\newcommand{\arcd}{\ifmmode^{\circ}\else$^{\circ}$\fi}
\newcommand{\arcm}{\ifmmode{'}\else$'$\fi}
\newcommand{\arcs}{\ifmmode{''}\else$''$\fi}
\newcommand{\MS}{{\rm M}\ifmmode_{\odot}\else$_{\odot}$\fi}
\newcommand{\RS}{{\rm R}\ifmmode_{\odot}\else$_{\odot}$\fi}
\newcommand{\LS}{{\rm L}\ifmmode_{\odot}\else$_{\odot}$\fi}

\newcommand{\Abstract}[2]{{\footnotesize\begin{center}ABSTRACT\end{center}
\vspace{1mm}\par#1\par
\noindent
{~}{\it #2}}}

\newcommand{\TabCap}[2]{\begin{center}\parbox[t]{#1}{\begin{center}
  \small {\spaceskip 2pt plus 1pt minus 1pt T a b l e}
  \refstepcounter{table}\thetable \\[2mm]
  \footnotesize #2 \end{center}}\end{center}}

\newcommand{\TableSep}[2]{\begin{table}[p]\vspace{#1}
\TabCap{#2}\end{table}}

\newcommand{\FigCap}[1]{\footnotesize\par\noindent Fig.\  %
  \refstepcounter{figure}\thefigure. #1\par}

\newcommand{\TableFont}{\footnotesize}
\newcommand{\TableFontIt}{\ttit}
\newcommand{\SetTableFont}[1]{\renewcommand{\TableFont}{#1}}

\newcommand{\MakeTable}[4]{\begin{table}[htb]\TabCap{#2}{#3}
  \begin{center} \TableFont \begin{tabular}{#1} #4 
  \end{tabular}\end{center}\end{table}}

\newcommand{\MakeTableSep}[4]{\begin{table}[p]\TabCap{#2}{#3}
  \begin{center} \TableFont \begin{tabular}{#1} #4 
  \end{tabular}\end{center}\end{table}}

\newenvironment{references}%
{
\footnotesize \frenchspacing
\renewcommand{\thesection}{}
\renewcommand{\in}{{\rm in }}
\renewcommand{\AA}{Astron.\ Astrophys.}
\newcommand{\AAS}{Astron.~Astrophys.~Suppl.~Ser.}
\newcommand{\ApJ}{Astrophys.\ J.}
\newcommand{\ApJS}{Astrophys.\ J.~Suppl.~Ser.}
\newcommand{\ApJL}{Astrophys.\ J.~Letters}
\newcommand{\AJ}{Astron.\ J.}
\newcommand{\IBVS}{IBVS}
\newcommand{\PASP}{P.A.S.P.}
\newcommand{\Acta}{Acta Astron.}
\newcommand{\MNRAS}{MNRAS}
\renewcommand{\and}{{\rm and }}
\section{{\rm REFERENCES}}
\sloppy \hyphenpenalty10000
\begin{list}{}{\leftmargin1cm\listparindent-1cm
\itemindent\listparindent\parsep0pt\itemsep0pt}}%
{\end{list}\vspace{2mm}}

\def\TYLDA{~}
\newlength{\DW}
\settowidth{\DW}{0}
\newcommand{\dw}{\hspace{\DW}}

\newcommand{\refitem}[5]{\item[]{#1} #2%
\def\REFARG{#3}\ifx\REFARG\TYLDA\else, {\it#3}\fi
\def\REFARG{#4}\ifx\REFARG\TYLDA\else, {\bf#4}\fi
\def\REFARG{#5}\ifx\REFARG\TYLDA\else, {#5}\fi.}

\newcommand{\Section}[1]{\section{#1}}
\newcommand{\Subsection}[1]{\subsection{#1}}
\newcommand{\Acknow}[1]{\par\vspace{5mm}{\bf Acknowledgments.} #1}
\pagestyle{myheadings}

\newfont{\bb}{ptmbi8t at 12pt}
\newcommand{\xrule}{\rule{0pt}{2.5ex}}
\newcommand{\xxrule}{\rule[-1.8ex]{0pt}{4.5ex}}
\def\thefootnote{\fnsymbol{footnote}}
\begin{center}
{\Large\bf The Optical Gravitational Lensing Experiment.\\
\vskip3pt
Catalog of high proper motion stars\\
\vskip6pt
towards the Magellanic Clouds\footnote{Based on observations obtained
with the 1.3~m Warsaw telescope at the Las Campanas Observatory of~the
Carnegie Institution of~Washington.}}
\vskip1cm
{\bf I.~~S~o~s~z~y~\'n~s~k~i$^1$,~~K.~~\.Z~e~b~r~u~\'n$^1$,~~A.~~U~d~a~l~s~k~i$^1$,\\
P.~R.~~W~o~\'z~n~i~a~k$^{2}$,~~M.~~S~z~y~m~a~{\'n}~s~k~i$^1$,~~M.~~K~u~b~i~a~k$^1$,\\
G.~~P~i~e~t~r~z~y~\'n~s~k~i$^{3,1}$,~~O.~~S~z~e~w~c~z~y~k$^1$,~~and~~\L.~~W~y~r~z~y~k~o~w~s~k~i$^1$}
\vskip3mm
$^1$Warsaw University Observatory, Al.~Ujazdowskie~4, 00-478~Warszawa, Poland\\
e-mail: (soszynsk,zebrun,udalski,msz,mk,pietrzyn,szewczyk,wyrzykow)@astrouw.edu.pl\\
$^2$ Los Alamos National Observatory, MS-D436, Los Alamos NM 85745, USA\\
e-mail: wozniak@lanl.gov\\
$^3$ Universidad de Concepci{\'o}n, Departamento de Fisica, Casilla 160--C, Concepci{\'o}n, Chile
\end{center}

\vspace*{4pt}
\Abstract{
We present a~Catalog of~high proper motion (HPM) stars detected in the
foreground of~central parts of~the Magellanic Clouds. The Catalog
contains 2161 objects in the 4.5~square degree area towards
the LMC, and 892 HPM stars in the 2.4~square degree area towards
the SMC. The Catalog is based on observations collected during four
years of~the OGLE-II microlensing survey.  The Difference Image Analysis (DIA)
of the images provided candidate HPM stars with proper motion as small as
4 mas/yr.  These appeared as pseudo-variables, and were all measured
astrometrically on all CCD images, providing typically about 400 data
points per star. The reference frame was defined by the majority
of~background stars, most of~them members of~the Magellanic Clouds.  The
reflex motion due to solar velocity with respect to the local standards
of~rest is clearly seen. The largest proper motion in our sample is
363~mas/yr. Parallaxes were measured with errors smaller than 20\% for
several stars.
}{}

\Section{Introduction}

Proper motions are of~considerable value for finding faint
nearby stars and studying overall gallactic structure.
Several groups undertook the proper motion surveys
searching for white dwarfs and other low luminosity stars in the
Galactic halo (e.g. Scholz \etal 2000, Wroblewski and
Costa 2001, Nelson \etal 2001, Oppenheimer \etal 2001). The proper
motions provided by the Hipparcos mission (Perryman \etal 1997) were
already used to study local Galactic structure (Dehnen and Binney 1998).

The majority of~hitherto existing HPM star catalogs are based on blinking
pairs of~photographic plates with several decades epoch
difference. For example New Luyten Two-Tenths (NLTT) Catalog (Luyten 1979,
Luyten and Hughes 1980), contains about 59 000 stars with proper motions
larger than 0.18 arcsec/yr.  It is the result of~comparing images taken at
epochs separated by more than thirty years. NLTT and the majority of~other
catalogs of~stars with proper motion contain objects found in uncrowded
sky regions, as it is difficult to recognize the same HPM star in a~crowded
environment.

Eyer and Wo\'zniak (2001) proposed a~new method of~finding HPM stars after a
relatively short observation period (several years) in very crowded
regions. The advantage of~this method is that HPM stars are detected
as a~by-product of~a~search for variable stars, i.e. no additional effort
is required. The method is based on the Difference Image Analysis (DIA) -- 
image subtraction algorithm developed by Alard and Lupton (1998) and Alard
(2000), and implemented by Wo\'zniak (2000). Further details of~the DIA
analysis of~the OGLE-II data may be found in papers by Wo{\'z}niak (2000)
and by {\.Z}ebru{\'n}, Soszy{\'n}ski and Wo{\'z}niak (2001). A~star which
changed its position relative to the reference frame defined by the majority
of background stars usually appears as a~pair of~variables separated by one
PSF (typically $ \sim 3 $ pixels).  One component of~the pair monotonically
increases its brightness, while the brightness of~the other component
decreases correspondingly, with the total flux remaining constant.

Eyer and Wo\'zniak (2001) proposed a~simple model that provides a~good
estimate of~the direction and the value of~stellar proper motion,
based on the measured rate of~flux variation and the
total flux of~a candidate for a~HPM star.
The value of~proper motion is given as

\begin{equation}
\mu = \frac{\sqrt{2\pi}\sigma}{F_{tot}}\gamma
\end{equation}
\noindent
where $\gamma$ is the slope of~the light curve, $F_{tot}$ is the total flux,
and $\sigma$ is the dispersion of~the PSF. The disadvantage of~this method
is a~fact that $F_{tot}$ often cannot be measured precisely.  Therefore,
it is important to perform astrometric measurements of~the candidate HPM star,
to have a~more robust determination of~its proper motion.  In other words,
the DIA software is very good in selecting HPM candidates, but it has to be
supplemented with astrometry.

In this paper we present the results of~a search for HPM stars, and the 
results of~astrometry in the data obtained by the OGLE-II
during four seasons of~observations of~the central parts of~the Magellanic
Clouds. We detected 2161 HPM stars towards the LMC and 892 stars towards
the SMC. We present coordinates, proper motions and
standard {\it BVI} photometry for all the HPM stars.
All data presented in this paper are available from the OGLE Internet
archive.

\vspace*{12pt}
\Section{Observational Data}

The observations were collected during the second phase of~the OGLE
experiment with the 1.3-m Warsaw telescope located at the Las Campanas
Observatory, Chile (operated by the Carnegie Institution
of~Washington). The telescope was equipped with the "first generation"
camera with a~SITe ${2048\times2048}$ CCD detector working in the
driftscan mode. The pixel size was 24~$\mu$m giving the 0.417 arcsec/pixel
scale. Observations of~the LMC were performed in the "slow" reading mode of
the CCD detector with the gain 3.8~e$^-$/ADU and readout noise about
5.4~e$^-$. Details of~the instrumentation setup can be found in Udalski,
Kubiak and Szyma\'nski (1997).

Regular observations of~the LMC fields started on January 6, 1997, while
observations of~the SMC started on June 26, 1997. About 4.5 square degrees
of~central parts of~the LMC (21 fields) and about 2.4 square degrees of~the
SMC (11 fields) were observed during four seasons. Data collected up to the
end of~May 2000 were used to detect HPM stars.

From 260 to 510 {\it I}-band observations were collected for each of~the
LMC fields, and about 300 {\it I}-band observations for each of~the SMC
fields. For each of~the LMC and SMC fields about 40 and 30 observations in
the {\it V} and {\it B}-band, respectively, were also collected. The
effective exposure time was 237, 173  and 125 seconds for {\it B}, {\it V}
and {\it I}-band, respectively. 
Median seeing of~the entire data set was about 1\zdot\arcs34. The analysis
was based on the {\it I}-band observations.

\Section{Selection of~HPM stars}

HPM stars appear in the difference images as a~bipolar flux residuals
changing approximately linearly over the entire period of
observations. Light curves of~the components of~the residual pair are
anti-correlated, because one of~them is created in the area where light of
star is reduced due to its motion, and the second is situated in
this region where the light increases. Axis of~the dipole fixes direction
of the star velocity vector in the sky. The members of~the pair are
separated roughly by $1-1.5$~arcsec, that is about 1 FWHM of~the PSF.
A~simple model of~the phenomenon was developed by Eyer and Wo\'zniak (2001).

The DIA software can detect HPM star by registering both residual poles, or,
depending on the adopted variability
thresholds, only one member of~the pair. In the second case 
a detection of~a HPM star is more uncertain, as the DIA light curve may
be due to genuine stellar variability.
Furthermore, the identification of~the DIA star on the reference
frame may be difficult in some cases, as our fields are very crowded.
In case of~a pair of~variables we searched for a~star exactly in the
middle of~the pair.  When only one component was
detected we searched for a~corresponding star on the reference frame
within 1~arcsec of~the DIA centroid.

Our selection of~HPM stars among the DIA variable stars
was divided into a~few phases.  First, we calculated a~correlation function
of light variations for each pair of~close DIA variables, and we easily
selected the pairs with anti-correlated variability and separated by
about $1-1.5$~arcsec. A~significant majority of~objects so selected
turned out to be HPM stars.

To select HPM stars in the case of~single monotonic variables
we examined the shapes of~their light curves. We
were looking for objects with linear variability during four observational
seasons by calculating correlations between light curves of~already detected
HPM stars and every remaining variable objects. Finally, we checked whether the
corresponding star could be found in the reference image in distance of
half FWHM (about $0.5-1$~arcsec).

Astrometric measurements were made for all candidates for HPM stars
chosen in both ways. Again we used programs of~DIA package. The large
2k$\times$8k frames were subdivided into 512$\times$128 pixel subframes, and
next subframes which contain a~given HPM star were re-sampled to
the pixel grid of~the OGLE template frame. When all of~the subframes
were in the same grid, we employed the DoPhot (Schechter, Mateo \& Saha
1995) to measure pixel coordinates of~the moving star. The last step was
transforming X-Y coordinates to the equatorial coordinates. We used
standard OGLE methods based in transformation derived with the
Digitized Sky Survey image (for details see Udalski \etal 1998).

\Section{Model fit}

Relative position of~a star with respect to the background stars may
change not only due to its proper motion. There are at least two
phenomena which change the centroid position of~stars with a~one-year
period. One of~them is the parallactic motion, which depends on the
distance to the star, the other is a~differential refraction offset --
second order effect of~atmospheric refraction, caused by a~difference
in star color compared to the average color of~reference stars. The
effect changes the apparent zenith angle of~the star and it is
proportional to $\tan(z)$.

The differential refraction is correlated with the time of~the year,
because the zenith angle of~observed regions is correlated with the time of
the year. At the beginning of~the observational season frames are taken
right above the eastern horizon, in mid-season most of~the observation
are being conducted at minimal possible values of~the zenith angle,
and in the end of~the season we observe again at large zenith angles
right above western horizon. The refraction always moves the star
toward the zenith, but according to the place in the sky, this is
different direction in the~equatorial coordinates. Dependence of
differential refraction on the time of~the year makes measuring
a~parallax of~the star very difficult.

Nevertheless we tried to reckon differential refraction and parallaxes
of the discovered HPM stars, because in several cases it was explicitly
seen, that yearly oscillations were superposed on the monotonic motion
in the sky (Fig. 1). To extract the parallax and the differential
refraction effects we prepared a~model of~these phenomena, and fitted
it to the measured coordinates of~the stars.

To obtain the dependence between the differential refraction
coefficient and the color of~the star we measured positions of~several
thousand stars from the LMC and the SMC. These objects are distant
enough to assume that proper motions and parallaxes are
undetectable. We used only bright ($I < 17$), rather isolated
stars. For every observational point of~the HPM star we computed the
zenith angle and the parallactic angle, and determined expected
differential refraction shift in RA and Dec directions. Next, using the
least square method, we fitted the value to the measured positions
of~the object.

Fig 2 presents the diagram of~color--differential refraction
coefficient for every analyzed HPM star. As expected there is a
clearly seen correlation between the color of~the star and value
of~differential refraction shift. We determined linear relationship
of~these values and for HPM stars with known (V-I) color index and we
corrected their coordinates for differential refraction.

The star's coordinates in the sky at the time $t$ can be described as follows:

\begin{equation}
\alpha = \mu_{\alpha} t + \pi \sin \gamma \sin \beta + r \tan z \sin p + \alpha_0
\end{equation}
\begin{equation}
\delta = \mu_{\delta} t + \pi \sin \gamma \cos \beta + r \tan z \cos p + \delta_0
\end{equation}
\noindent
where $\alpha$ and $\delta$ are respectively right ascension and
declination of~the star, $\pi$ is the parallax, $r$ is the differential
refraction coefficient, $\gamma$ is an~angular distance to the Sun in
the celestial sphere, $z$ is a~zenith angle, $\beta$ and $p$ are
angles in the celestial sphere between direction of~respectively
parallax shift and refractional shift and direction to the celestial
North pole, finally $\alpha_0$ and $\delta_0$ are constant.

Knowing time $t$ of~every observation we computed values of~$\gamma$,
$z$, $\beta$ and $p$. The position of~the Sun was computed using
formulae for the Sun position by van Flandern and Pulkkinen
(1979). The differential refraction coefficient $r$ was estimated
using linear $r$--$(V-I)$ dependence. Next, using least square method
we obtained values of~$\mu_{\alpha}$, $\mu_{\delta}$, $\pi$,
$\alpha_0$ and $\delta_0$.

\Section{Results}

Final list of~HPM stars consists of~2161 objects found towards the LMC
and 892 stars found in a~foreground of~the SMC. Using our method we
could measure proper motions as small as 4 mas/yr. We note, that the
method is insensitive to proper motions larger than $\sim 1$
arcsec/yr, because light curves of~the components of~the residual pair
would quickly become constant.

Because the large number of~objects in the catalog decided to make data available only in the electronic form. The lists of~the HPM stars can be accessed {\it via} anonymous ftp at the following addresses:

~~~~~{\it ftp://bulge.princeton.edu/ogle/ogle2/hpm/}

~~~~~{\it ftp://sirius.astrouw.edu.pl/ogle/ogle2/hpm/}

The ASCII files contain equatorial coordinates, measured proper motions, {\it
I}-band photometry, $V-I$ and $B-V$ colors of~all discovered HPM stars
in the lines of~sight to the LMC and the SMC. The tables contain more 
lines than the number of~discovered HPM stars, because 55 objects towards 
the LMC, and 10 stars towards the SMC were detected twice -- at the 
overlapping parts of~adjacent fields.

The WWW interface of~the catalog is available at the addresses:

~~~~~{\it http://bulge.princeton.edu/$\sim$ogle/ogle2/hpm/}

~~~~~{\it http://www.astrouw.edu.pl/$\sim$ogle/ogle2/hpm/}

Apart from basic information about the stars, one can see here finding charts for every objects in the catalog and diagrams with the astrometric measurements of~the stars.

In Figs. 3 and 4 we present location and motion vectors of~all discovered
HPM stars and contours of~the observed fields. It can be seen that in a
foreground of~fields covering central part of~the LMC (fields
LMC\_SC2--LMC\_SC7) we discovered on average more HPM stars than
towards the remaining LMC fields. This inhomogeneous density
of~detected HPM stars can be explained by different number and
different period of~observations. OGLE-II had collected about 500 {\it
I}-band observations of~fields LMC\_SC2--LMC\_SC7, and about 250-350
observations for fields covering regions outside strict center of~the
LMC. Regular observations of~these fields started about six month
after LMC\_SC2--LMC\_SC7. Shorter time baseline used for analysis
probably caused that on the variability maps, created by DIA, some HPM
stars remained below thresholds.

Color magnitude diagrams of~the HPM stars from the catalog are presented in
Fig. 5. Tiny dots indicates CMD diagrams of~the background LMC and SMC stars.
In Fig. 6 we present reduced proper motion diagram for our HPM
stars. Reduced proper motion ($m+5\log\mu$) is a rough approximation of
the absolute magnitude ($M=m+5\log\pi+5$) up to a zero point offset.

Figs. 7 and 8 presents two-dimensional distribution of~the proper motion
vectors from the catalog. It is clearly seen, that stars prefer moving
in the specified directions: roughly to the North in the~foreground
of~the LMC and roughly to the East in the SMC regions. Average proper
motion of~the stars is more than 20~mas/yr, so it could not be an
effect of~proper motion of~the Magellanic Clouds, because it is less
than 2 mas/yr. Probably it is an~effect of~Solar movement relative to
the local standard of~rest. Simple calculations confirm this
hypothesis. Assuming that the Sun is moving toward Hercules
constellation ($\alpha=18^h$, $\delta=+30\arcd$), nearby stars in 
the foreground of~the LMC and SMC should move in the sky
in the directions indicated with the arrows in Figs. 7 and 8.

In Table~1 we present parallaxes, positions, proper motions and magitudes
of~stars, for which the parallaxes were measured with errors smaller than
20\%. We found 25 such stars in the foreground of~the LMC and 13 towards
the SMC.

\renewcommand{\arraystretch}{1.2}
\renewcommand{\TableFont}{\scriptsize} 
\MakeTableSep{l@{\hspace{2pt}}
c@{\hspace{5pt}}c@{\hspace{5pt}}c@{\hspace{6pt}}r@{\hspace{5pt}}r@{\hspace{4pt}}r@{\hspace{4pt}}c@{\hspace{4pt}}c@{\hspace{4pt}}c@{\hspace{4pt}}}{11.5cm}
{Stars with measurable parallaxes. Errors of~the paralaxes are smaller than 20\%.}{\hline
\noalign{\vskip4pt} Field & Star ID   & RA (J2000)  & DEC (J2000) &   $\pi$~~~& $\mu_{\alpha^*}$~~~~& $\mu_{\delta}$~~~~~& $I$    & $V-I$ & $B-V$ \\
           &             &             &             & [mas]   & [mas/yr]
           & [mas/yr]     & [mag]  & [mag] & [mag]  \\ 
\noalign{\vskip4pt}
\hline
\noalign{\vskip4pt}
LMC\_SC1 & 2000-2001 & $5\uph33\upm26\zdot\ups28$ & $-69\arcd48\arcm26\zdot\arcs8$ &  9~~ & 41~~~~ & 65~~~~ & 15.923 & 2.844 & 1.664 \\
LMC\_SC2 & 273-274 & $5\uph31\upm02\zdot\ups19$ & $-70\arcd16\arcm43\zdot\arcs0$ &  9~~ & -7~~~~ & -70~~~~ & 15.304 & 2.489 & 2.173 \\
LMC\_SC3 & 1246-0 & $5\uph28\upm41\zdot\ups98$ & $-69\arcd59\arcm42\zdot\arcs1$ &  9~~ & -36~~~~ & -3~~~~ & 17.114 & 3.779 & -- \\
LMC\_SC3 & 3350-0 & $5\uph28\upm30\zdot\ups58$ & $-69\arcd37\arcm29\zdot\arcs0$ &  8~~ & -5~~~~ & -18~~~~ & 16.300 & 3.239 & 1.139 \\
LMC\_SC4 & 2134-0 & $5\uph26\upm45\zdot\ups90$ & $-69\arcd51\arcm04\zdot\arcs0$ & 10~~ & -28~~~~ & -81~~~~ & 14.701 & 2.330 & 1.303 \\
LMC\_SC4 & 3877-0 & $5\uph25\upm45\zdot\ups02$ & $-69\arcd30\arcm29\zdot\arcs8$ &  9~~ & 45~~~~ & -93~~~~ & 15.318 & 2.455 & 1.453 \\
LMC\_SC4 & 4246-4247 & $5\uph27\upm07\zdot\ups14$ & $-69\arcd26\arcm37\zdot\arcs6$ &  9~~ & -1~~~~ & -22~~~~ & 16.101 & 2.855 & 1.574 \\
LMC\_SC4 & 4539-4541 & $5\uph27\upm31\zdot\ups33$ & $-69\arcd22\arcm09\zdot\arcs6$ & 16~~ & 79~~~~ & 134~~~~ & 14.718 & 3.718 & 1.383 \\
LMC\_SC6 & 3491-3492 & $5\uph20\upm43\zdot\ups13$ & $-69\arcd22\arcm46\zdot\arcs6$ & 12~~ & 32~~~~ & 118~~~~ & 16.607 & 3.484 & -- \\
LMC\_SC6 & 4307-0 & $5\uph22\upm33\zdot\ups42$ & $-69\arcd12\arcm02\zdot\arcs6$ &  8~~ & -32~~~~ & 16~~~~ & 13.892 & 2.383 & 1.645 \\
LMC\_SC7 & 429-0 & $5\uph18\upm58\zdot\ups28$ & $-69\arcd47\arcm47\zdot\arcs5$ &  9~~ & 67~~~~ & 113~~~~ & 13.342 & 2.484 & 1.341 \\
LMC\_SC8 & 532-533 & $5\uph17\upm26\zdot\ups10$ & $-69\arcd41\arcm21\zdot\arcs2$ & 17~~ & 106~~~~ & 155~~~~ & 14.794 & 3.569 & 1.152 \\
LMC\_SC8 & 1411-1413 & $5\uph16\upm44\zdot\ups98$ & $-69\arcd29\arcm30\zdot\arcs3$ & 19~~ & -131~~~~ & 217~~~~ & 13.574 & 3.515 & 1.563 \\
LMC\_SC9 & 1361-1362 & $5\uph12\upm49\zdot\ups43$ & $-69\arcd18\arcm37\zdot\arcs6$ &  7~~ & 78~~~~ & 8~~~~ & 13.926 & 2.415 & 1.579 \\
LMC\_SC9 & 2148-0 & $5\uph13\upm23\zdot\ups89$ & $-69\arcd06\arcm02\zdot\arcs2$ & 25~~ & -34~~~~ & 46~~~~ & 16.366 & 4.478 & -- \\
LMC\_SC9 & 2497-2498 & $5\uph12\upm40\zdot\ups08$ & $-68\arcd58\arcm30\zdot\arcs3$ & 22~~ & -45~~~~ & 75~~~~ & 15.489 & 3.595 & 1.333 \\
LMC\_SC12 & 1245-1246 & $5\uph05\upm56\zdot\ups07$ & $-69\arcd22\arcm59\zdot\arcs9$ &  9~~ & 13~~~~ & -30~~~~ & 13.438 & 2.476 & 1.433 \\
LMC\_SC12 & 1367-1368 & $5\uph05\upm44\zdot\ups54$ & $-69\arcd19\arcm21\zdot\arcs7$ & 11~~ & 27~~~~ & -7~~~~ & 14.205 & 2.606 & 1.432 \\
LMC\_SC13 & 351-352 & $5\uph05\upm26\zdot\ups99$ & $-69\arcd04\arcm54\zdot\arcs8$ & 20~~ & 108~~~~ & 99~~~~ & 14.233 & 3.409 & -- \\
LMC\_SC13 & 2914-2915 & $5\uph07\upm04\zdot\ups47$ & $-68\arcd18\arcm31\zdot\arcs4$ & 10~~ & -35~~~~ & -82~~~~ & 14.165 & 2.797 & 1.413 \\
LMC\_SC16 & 700-701 & $5\uph35\upm19\zdot\ups34$ & $-70\arcd15\arcm23\zdot\arcs8$ & 14~~ & -9~~~~ & -74~~~~ & 14.432 & 2.756 & 1.538 \\
LMC\_SC16 & 1263-1264 & $5\uph35\upm41\zdot\ups03$ & $-70\arcd03\arcm31\zdot\arcs5$ & 10~~ & -7~~~~ & 31~~~~ & 13.979 & 2.478 & 1.554 \\
LMC\_SC21 & 592-593 & $5\uph20\upm07\zdot\ups30$ & $-70\arcd35\arcm31\zdot\arcs8$ & 25~~ & -44~~~~ & 242~~~~ & 15.204 & 3.384 & 1.864 \\
LMC\_SC21 & 667-0 & $5\uph20\upm16\zdot\ups60$ & $-70\arcd32\arcm40\zdot\arcs6$ & 16~~ & -14~~~~ & 40~~~~ & 15.213 & 3.170 & 1.857 \\
LMC\_SC21 & 1241-0 & $5\uph21\upm05\zdot\ups67$ & $-70\arcd12\arcm42\zdot\arcs0$ & 16~~ & 55~~~~ & 34~~~~ & 13.980 & 2.346 & 1.822 \\
SMC\_SC3 & 210-211 & $0\uph42\upm50\zdot\ups17$ & $-73\arcd29\arcm19\zdot\arcs6$ & 11~~ & 48~~~~ & -93~~~~ & 13.784 & 2.189 & 1.625 \\
SMC\_SC4 & 1681-0 & $0\uph46\upm12\zdot\ups77$ & $-72\arcd50\arcm25\zdot\arcs1$ & 12~~ & -51~~~~ & -23~~~~ & 13.729 & 2.585 & -- \\
SMC\_SC5 & 370-0 & $0\uph49\upm00\zdot\ups99$ & $-73\arcd26\arcm06\zdot\arcs9$ & 12~~ & 75~~~~ & 44~~~~ & 15.408 & 2.822 & 1.732 \\
SMC\_SC7 & 61-62 & $0\uph55\upm59\zdot\ups74$ & $-73\arcd19\arcm19\zdot\arcs0$ & 15~~ & -142~~~~ & -57~~~~ & 14.985 & 2.735 & 1.865 \\
SMC\_SC7 & 759-0 & $0\uph55\upm38\zdot\ups47$ & $-72\arcd58\arcm01\zdot\arcs5$ & 12~~ & 40~~~~ & 13~~~~ & 14.692 & 2.459 & 1.523 \\
SMC\_SC7 & 1403-0 & $0\uph54\upm51\zdot\ups69$ & $-72\arcd38\arcm24\zdot\arcs4$ & 15~~ & 47~~~~ & -46~~~~ & 14.175 & 2.647 & 1.638 \\
SMC\_SC7 & 1539-1540 & $0\uph55\upm56\zdot\ups30$ & $-72\arcd34\arcm34\zdot\arcs8$ & 13~~ & 37~~~~ & -24~~~~ & 13.812 & 2.780 & 1.549 \\
SMC\_SC8 & 934-935 & $1\uph00\upm10\zdot\ups42$ & $-72\arcd30\arcm14\zdot\arcs2$ &  8~~ & -158~~~~ & -155~~~~ & 15.021 & 2.467 & 1.798 \\
SMC\_SC10 & 375-376 & $1\uph04\upm49\zdot\ups97$ & $-72\arcd28\arcm57\zdot\arcs8$ & 22~~ & 352~~~~ & -90~~~~ & 15.461 & 3.195 & 1.833 \\
SMC\_SC10 & 631-632 & $1\uph03\upm56\zdot\ups50$ & $-72\arcd05\arcm57\zdot\arcs1$ & 17~~ & 59~~~~ & -16~~~~ & 14.987 & 2.923 & 1.725 \\
SMC\_SC11 & 156-0 & $1\uph06\upm39\zdot\ups91$ & $-72\arcd55\arcm07\zdot\arcs2$ &  9~~ & -27~~~~ & -25~~~~ & 16.362 & 2.973 & -- \\
SMC\_SC11 & 551-552 & $1\uph09\upm09\zdot\ups19$ & $-72\arcd35\arcm43\zdot\arcs4$ &  8~~ & 86~~~~ & 12~~~~ & 15.616 & 2.765 & 1.533 \\
SMC\_SC11 & 780-0 & $1\uph08\upm10\zdot\ups29$ & $-72\arcd25\arcm03\zdot\arcs4$ & 15~~ & 86~~~~ & 8~~~~ & 16.895 & 3.809 & -- \\
\hline
$\mu_{\alpha^*}=\mu_{\alpha \cos\delta}$
}

\Section{Completeness of~the Catalog}

One should note that we discovered the HPM stars using methods worked
out for variable stars detection, not for searching fast moving
objects. Therefore a~completeness of~the HPM catalog strictly depends on
an efficiency of~these algorithms. The DIA programs are finding
variable objects with the PSF similar to the PSF of~stars. Shapes
of~each member of~the HPM residual dipole are only roughly similar to
the PSF of~stars, so we know for certain, that part of~the HPM stars
in our fields were not detected. The fact that only less than half~of~HPM
candidates had both poles of~a pseudo variable pair
detected with the DIA implies that there were some HPM stars 
for which neither pole was detected, and the object was missed.

In Fig. 9 we present {\it I}-band magnitudes plots against the
proper motions. It is clearly visible, that there is a~minimal value
of~proper motion, below which star's movement cannot be detected using
our methods. We registered proper motions as small as 4 mas/yr for
stars of~$I=14\div16$ mag. For brighter and especially fainter stars the
detection limit is higher.

Fig. 3 shows that the completeness of~the catalog strongly depends
on a~number and time-span of~observations. Density of~the HPM stars, which
should be independent of~background LMC stars, is significantly higher
in more frequently observed regions. Therefore we consider that the
completeness of~the HPM sample is higher in the fields
LMC\_SC2--LMC\_SC7 than in other fields.

Recently Alcock \etal (2001) detected 154 HPM stars along the lines
of sight of~the Galactic Bulge and the Large Magellanic Cloud, selecting 
them from the MACHO catalog of~variable stars. The catalog contains stars
with proper motion larger than 30 mas/yr. Towards the LMC, 64 HPM
stars were detected, 20 of~which are in the line of~sight of~OGLE
fields. In the SMC MACHO collaboration detected one HPM star. We
conducted cross identification of~the stars in both catalogs, during
which we found out that 19 HPM stars were rediscovered (18 stars in the
LMC and 1 star in the SMC). One of~not detected stars was too bright
($I=12.4$) and stayed just above our saturation level. The other star has
been missed, because correlation coefficient between shapes of~the
residual poles and the PSF of~the stars was too low. 

We obtained about 80\% completness of~the stars with $\mu>30$ mas/yr and
$I<17$ mag by comparing the HPM stars in the
overlapping regions in the neighboring fields. One should note that
regions at the field edges are biased by smaller number
of~observations, due to imperfections in telescope pointing, what
reduces the completeness of~the catalog in these areas. In total 33
stars with $\mu>30$ mas/yr should be paired with counterparts in the
overlapping fields. We found counterparts in 27 cases. When we compared
subsample of~stars with proper motion larger than 50 mas/yr, we
identified all counterparts in the neighboring fields. The completeness
of~the catalog is decreasing for fainter stars and with smaller proper
motion. We estimate that for objects with proper motion smaller than
10~mas/yr, the completeness is less than 50\%.

\Section{Conclusions}

Difference Image Analysis offers a~great possibility of~detecting HPM
stars in dense stellar regions. The moving stars are discovered in the
procedure of~variable star detection, so the process does not require
any special efforts.

The knowledge of~stars in the Solar neighborhood is the starting
point for projects of~determining stellar luminosity and mass
functions and other properties of~stars. Large number of~HPM stars in
this catalog will be useful to study Galactic dynamics. Many of~the
objects are undoubtedly low-luminosity stars, such as M-dwarfs. These
stars are hot topics, because their features are still a~matter of
debate.

HPM stars in a~foreground of~densely crowded stellar fields offer
possibility for predicting lensing events many years in
advance. Paczy\'nski (1996) showed that astrometric effect
of~gravitational microlensing, which can be measured by Hubble Space
Telescope, may enable a~direct mass determination for the neighboring
stars. The cross-section for astrometric effect is larger than the
cross-section for photometric effect of~gravitational microlensing
measurable from the ground. It means that predictions of~microlensing
by stars from the HPM catalogs may significantly increase the number
of targets for astrometric measurements from space.

The Space Interferometry Mission (SIM) telescope, planned to be
launched in 2009, will allow to measure positions in the sky with
about 4 $\mu$as accuracy. It could permit the determination of~the
distance, mass and radius of~the lensing object (Paczy\'nski
1998). Gould (2000) and Salim \& Gould (2000) discussed the selection of
candidates for SIM observations.

Third phase of~OGLE project (OGLE-III) started in June, 2001. The
telescope has been equipped with a~new generation CCD mosaic camera
$8192\times8192$ pixels, what considerably increased quality of~the
images. We expect that OGLE-III data will allow to detect proper
motions as small as 1-2 mas/yr, and to determine exact parallaxes of
the neighboring stars.

\Acknow{We would like to thank Prof. Bohdan Paczy\'nski for many
discussions and support in this work. This work was partly supported by the
KBN grant 5P03D02520 to I. Soszy\'nski and 2P03D01418 to M.~Kubiak. Partial
support was also provided by the NSF grant AST-9830314 to
B.~Paczy\'nski. Support for PW was provided by the Laboratory
Directed Research and Development funds at LANL.}

\begin{figure}[p]
\hglue-6mm
\includegraphics[bb=40 30 550 730,width=13cm]{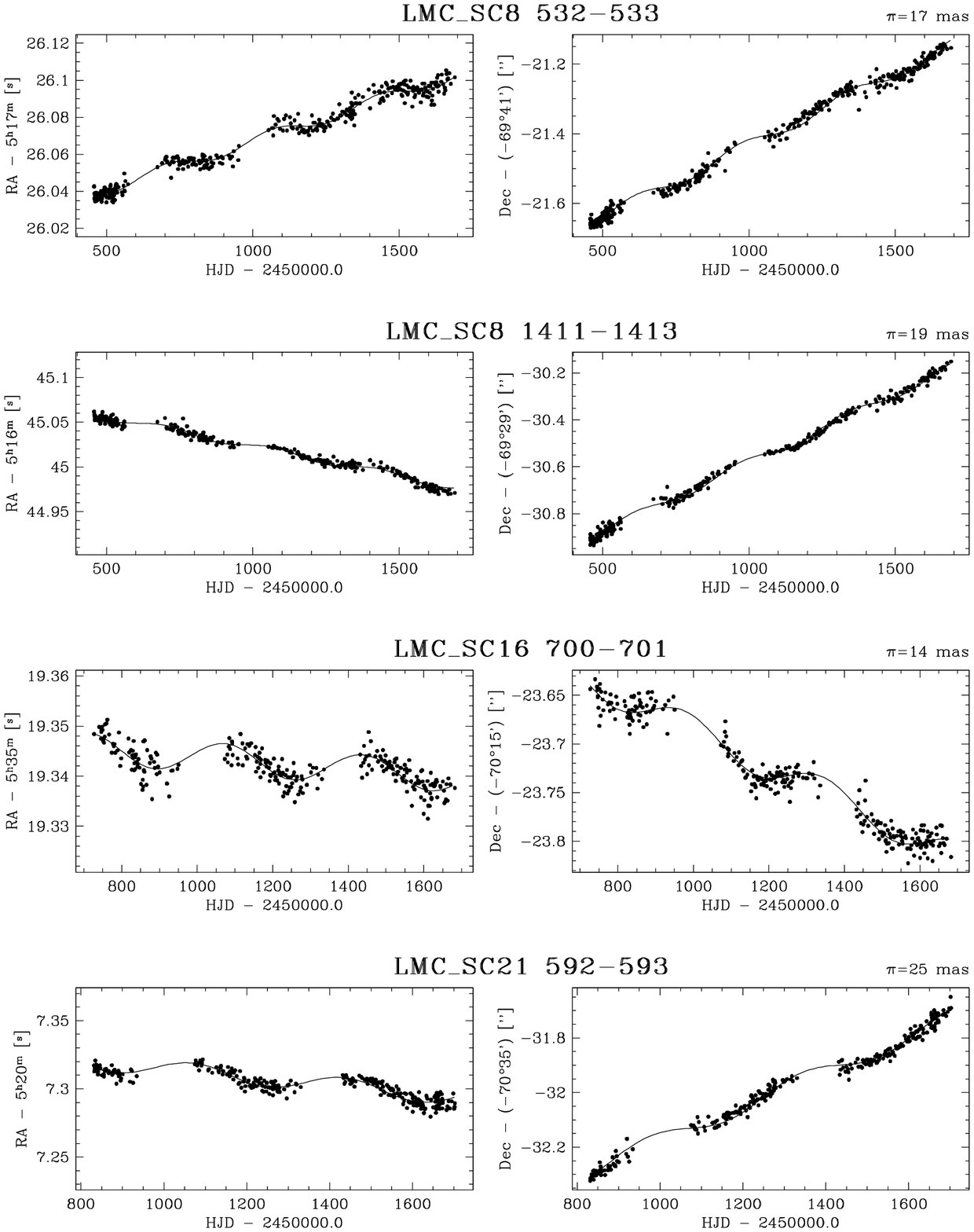}
\vskip3pt
\FigCap{Equatorial coordinates of~the four HPM stars from the LMC with the model fit. The differential refraction shift is subtracted from every point.}
\end{figure}

\begin{figure}[p]
\hglue-6mm
\vspace{16cm}
\vskip3pt
\FigCap{Correlation between the colors of~stars and the fitted differential
refraction coefficient for stars from the LMC (upper panel) and from
the SMC (lower panel). The dispersion of~these diagrams is an~effect
of local differences in background stars' color and uncertainties of
the astrometric measurements.}
\end{figure}

\begin{figure}[p]
\hglue-6mm
\includegraphics[bb=40 60 550 730,width=13cm]{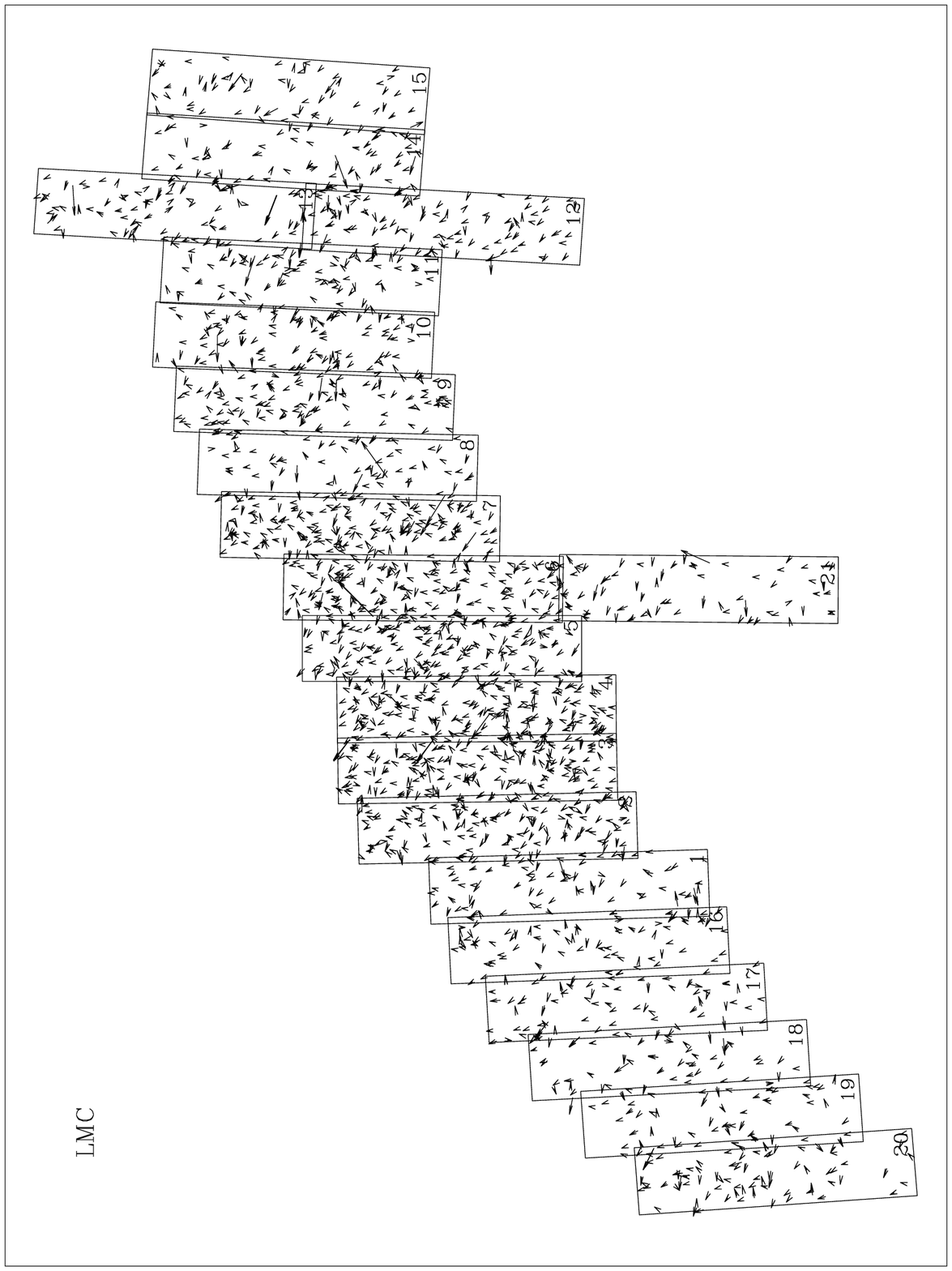}
\vskip3pt
\FigCap{Map of~the LMC fields with discovered HPM stars.}
\end{figure}

\begin{figure}[p]
\hglue-6mm
\includegraphics[bb=40 30 550 730,width=13cm]{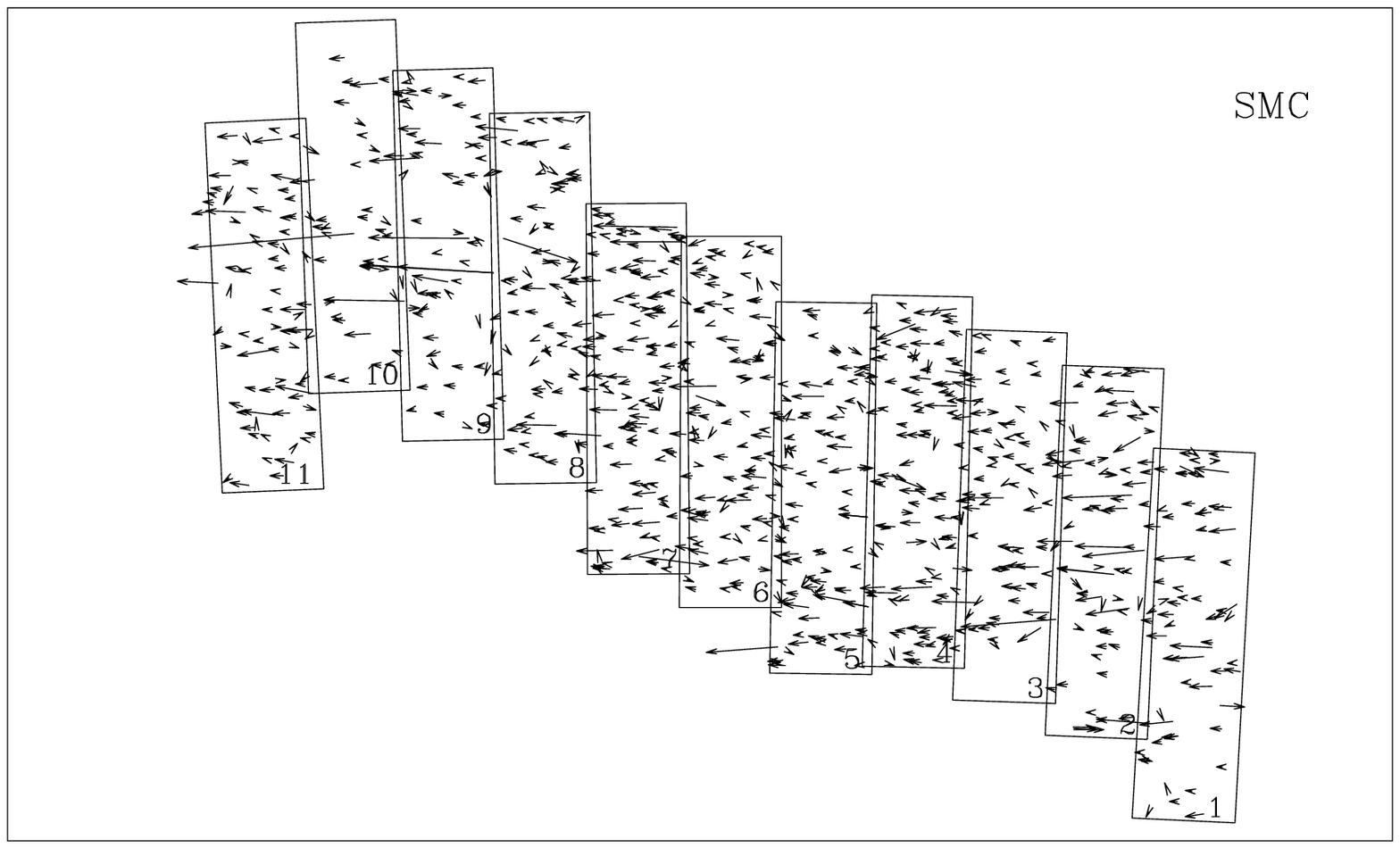}
\vskip3pt
\FigCap{Map of~the SMC fields with discovered HPM stars.}
\end{figure}

\begin{figure}[p]
\hglue-6mm
\vspace{16cm}
\vskip3pt
\FigCap{Color-Magnitude diagram of~the HPM stars observed
towards the LMC (upper panel) and the SMC (lower panel). Tiny dots in the
upper and lower panel indicate about 50 000 stars from LMC\_SC3 and
SMC\_SC1 fields respectively}
\end{figure}

\begin{figure}[p]
\hglue-6mm
\includegraphics[bb=40 55 550 730,width=13cm]{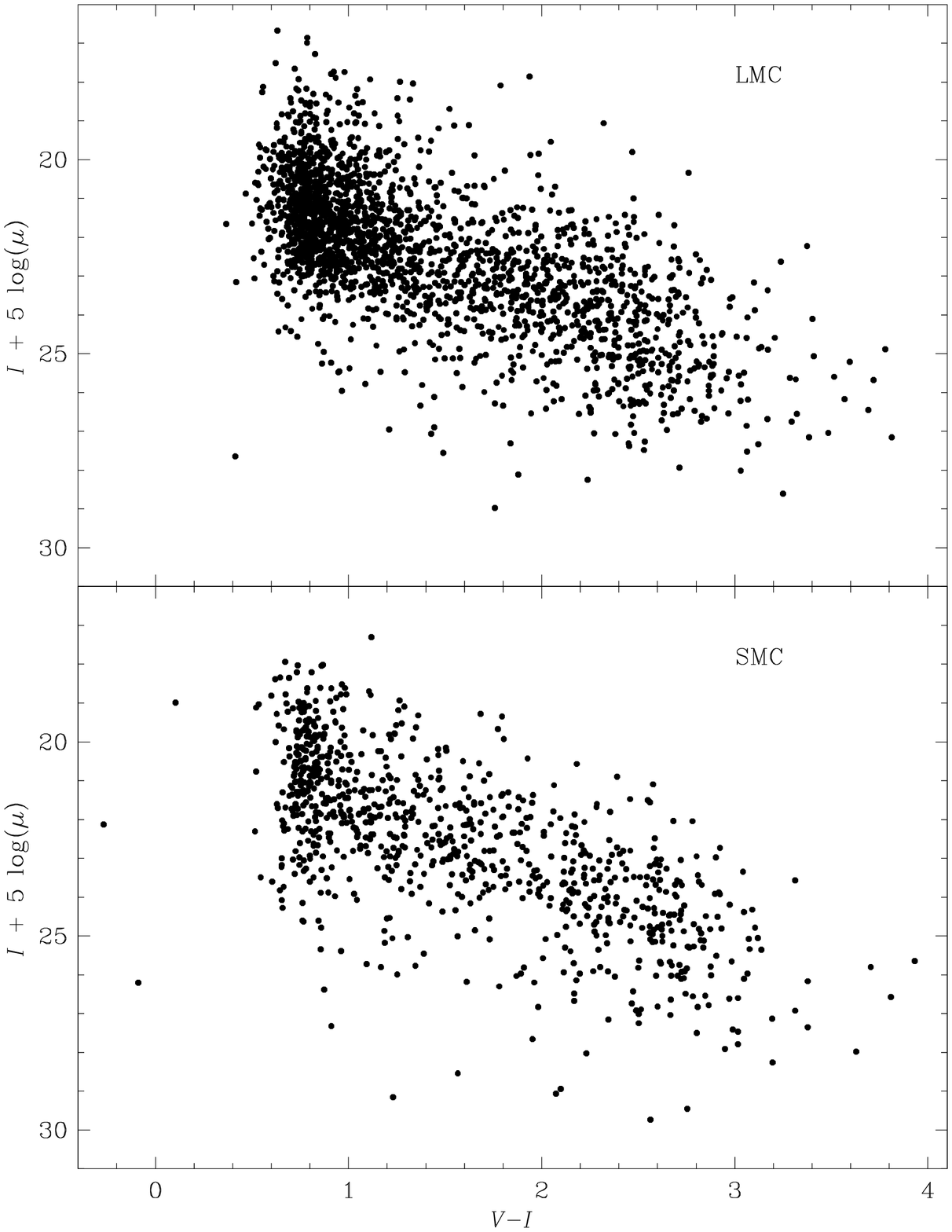}
\vskip3pt
\FigCap{Reduced proper motion diagram for HPM stars observed towards the LMC (upper panel) and the SMC (lower panel).}
\end{figure}

\begin{figure}[p]
\hglue-6mm
\includegraphics[bb=40 215 550 730,width=13cm]{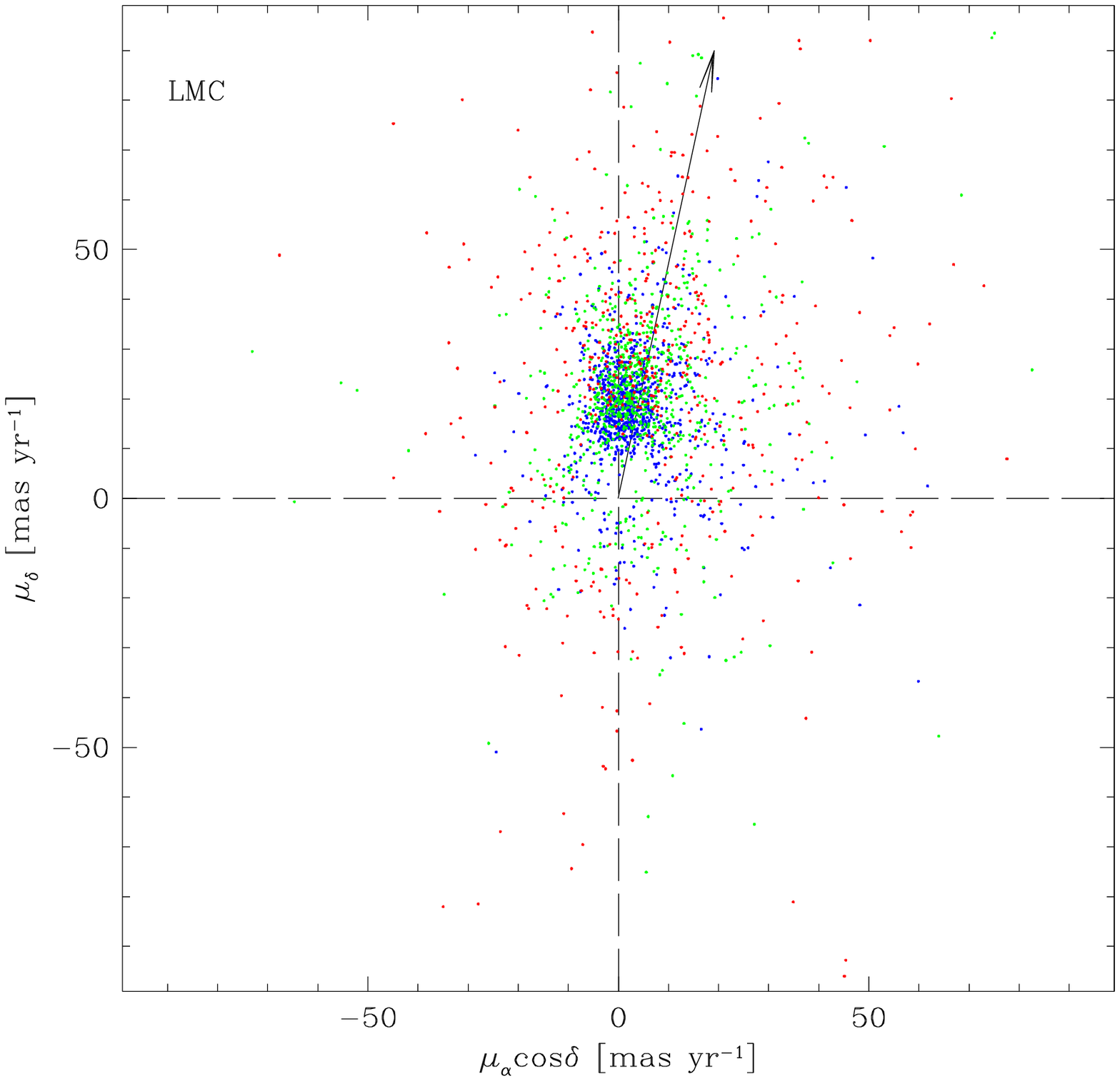}
\vskip3pt
\FigCap{Two-dimensional distribution of~proper motions $\mu_{\alpha}$,
$\mu_{\delta}$ of~stars towards the LMC. The long arrow shows the direction
opposite to the solar apex. Colors of the points indicate colors of the
stars -- blue: $(V-I)<1$, green: $1<(V-I)<2$ and red: $(V-I)>2$.}
\end{figure}

\begin{figure}[p]
\hglue-6mm
\includegraphics[bb=40 215 550 730,width=13cm]{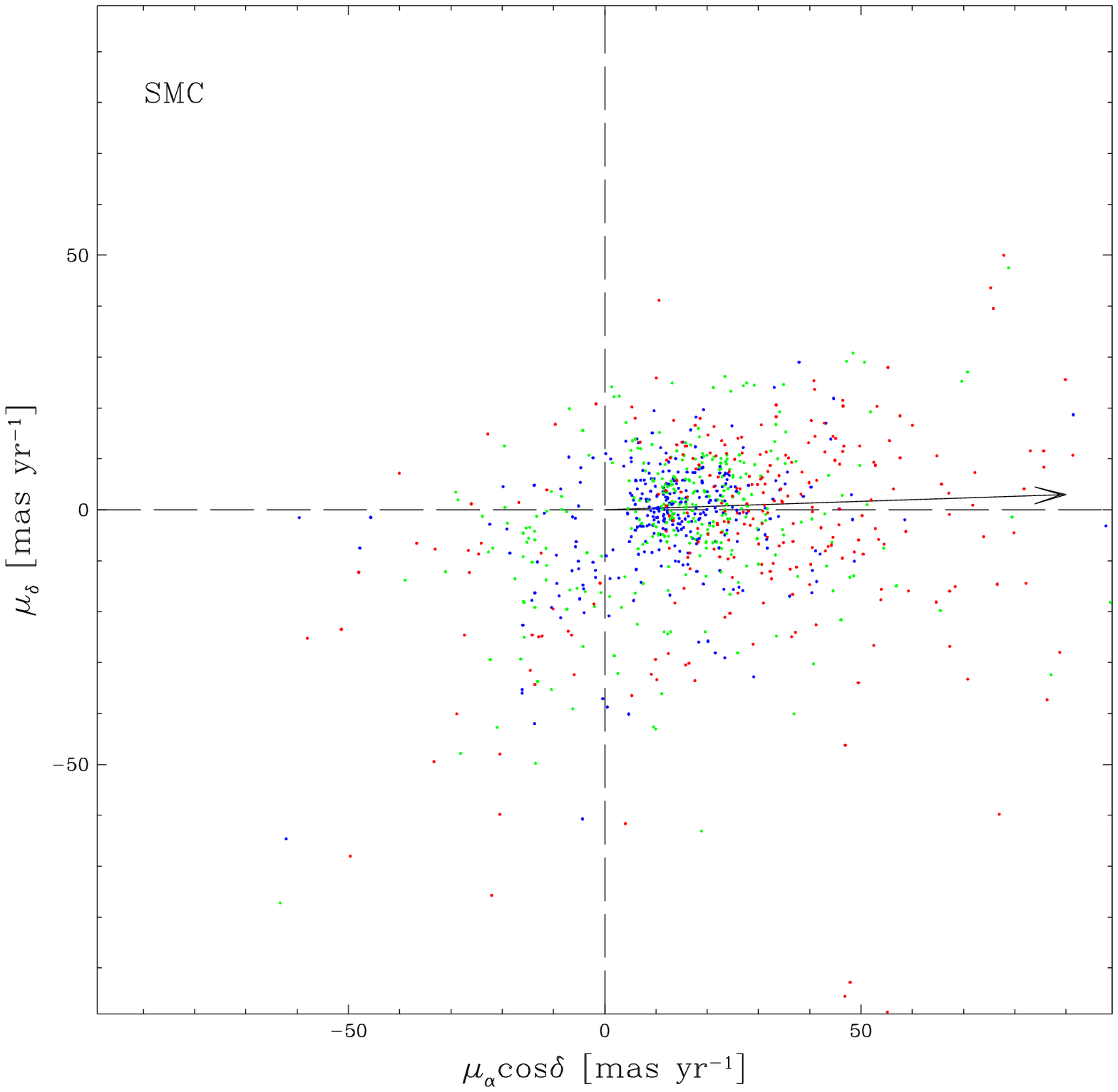}
\vskip3pt
\FigCap{Two-dimensional distribution of~proper motions $\mu_{\alpha}$,
$\mu_{\delta}$ of~stars towards the SMC. The long arrow shows the direction
opposite to the solar apex. Colors of the points were discribed in the caption to Fig. 7}
\end{figure}

\begin{figure}[p]
\hglue-6mm
\includegraphics[bb=40 45 550 730,width=13cm]{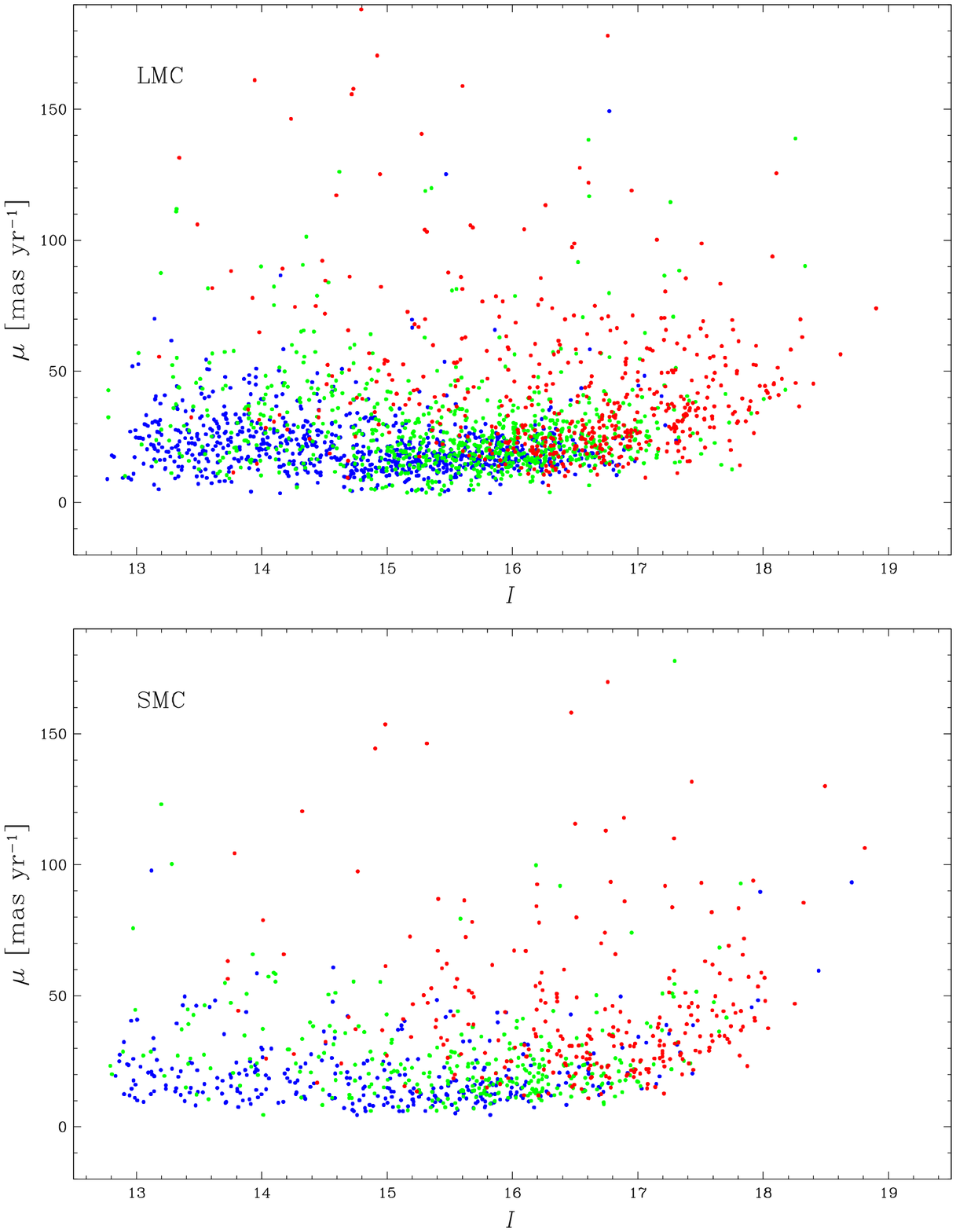}
\vskip3pt
\FigCap{Luminosity versus proper motion diagram of~the HPM stars observed
towards the LMC (upper panel) and the SMC (lower panel). One can notice dependence between minimal measured value of~proper motions and magnitude of~stars. Colors of the points were discribed in the caption to Fig. 7}
\end{figure}

\end{document}